\documentclass[preprint,12pt]{elsarticle}
\usepackage{amsmath}
\usepackage{amssymb}
\usepackage[pdftex,colorlinks]{hyperref}
\usepackage{multirow}

\biboptions{compress}

\begin{document}

\begin{frontmatter}

\title{Distributed quantum election scheme}

%% use optional labels to link authors explicitly to addresses:
\author{Rui-Rui Zhou}
\author{Li Yang\corref{1}}%\ead{yangli@iie.ac.cn}
\cortext[1]{Corresponding author email: yangli@iie.ac.cn}
\address{State Key Laboratory of Information Security, Institute of Information Engineering, Chinese Academy of Sciences, Beijing 100195, China}
%% \address[label2]{<address>}

\begin{abstract}
In an electronic voting protocol, a distributed scheme can be used for forbidding the malicious acts of the voting administrator and the counter during the election, but it cannot prevent them from collaborating to trace the ballots and destroy their privacy after the election. We present a distributed anonymous quantum key distribution scheme and further construct a distributed quantum election scheme with a voting administrator made up of more than one part. This quantum election scheme can resist the malicious acts of the voting administrator and the counter after the election and can work in a system with lossy and noisy quantum channels.

\end{abstract}

\begin{keyword}
quantum election \sep distributed scheme \sep conjugate coding
%% keywords here, in the form: keyword \sep keyword

%% MSC codes here, in the form: \MSC code \sep code
%% or \MSC[2008] code \sep code (2000 is the default)

\end{keyword}

\end{frontmatter}

%%
%% Start line numbering here if you want
%%
% \linenumbers

%% main text
\section{Introduction}

In a large-scale election, the problem that most concerns the voters is the privacy of the election. That is, an eligible voter does not want anybody to track his/her ballot at any time. Further, eligibility and unreusability are other serious problems. That is, only eligible voters are permitted to vote, and each eligible voter can vote successfully only once. In view of the properties described in earlier papers$^{[1-3]}$, an ideal election scheme should have the following properties: completeness, soundness, privacy, eligibility, unreusability, fairness and verifiability.  In order to ensure these properties, considerable attention has been paid to election schemes. These schemes can be divided to two parts: electronic election and quantum election. The homomorphic-encryption-based scheme$^{[4-7]}$, mix-net-based scheme$^{[1,8]}$, and blind signature-based scheme$^{[2,3,9,10]}$ are there main types of electronic voting schemes. These schemes can efficiently solve the drawback of achieving privacy and fairness at the same time. However, the security of these schemes is based on the difficulty of solving certain mathematical problems; it will be threatened by the use of a quantum computer. When a quantum computer is used, mathematical security can no longer prevent the attacker from knowing whom a voter voted for in the election.

Quantum cryptography can be used for solving the problem of unconditional security and privacy that mentioned in the earlier electronic voting schemes, which use the fundamental laws of quantum physics to ensure unconditional security$^{[11-15]}$. For example, we can use the quantum non-cloning theorem for unconditionally secure quantum key distribution$^{[16-19]}$, and we can use quantum anonymous transmission$^{[20]}$ to conceal the identity of the sender of the messages. Considerable attention has been paid to quantum election protocols. Vaccaro et al. propose a quantum election protocol$^{[21]}$ by adding different local quantum operations to an entangled quantum state that distributed over separated sites in 2007, the physical inaccessibility of any one site is sufficient to guarantee the anonymity of the votes; Hillery et al. presents a similar protocol$^{[22]}$, in which the initial state of the system is denoted by a quantum state $|\Omega_0\rangle$, and the eligible voter $V_j$ expresses his choice through different operations $U_k^j$ to the initial quantum state(the value of $k$ depends on his choice). The counter extracts the outcome of the election through a complicated measurement of the final quantum state and cannot traces the ballot of a specific voter. These protocols are relatively significant progress in the field of quantum elections. On the one hand, these protocols efficiently guarantee the security of the election and ensure the anonymity of the voters; however, on the other hand, as the outcome of the measurement is the number statistics of all the votes, no voter can trace his ballot from the outcome, and thus, he cannot ensure whether he has voted successfully or not. Further, the reading of the outcome statistics is a complicated measurement: there is no reliable way of knowing whether a voter has voted more than once, and the voting for the candidates of the election is restricted to yes or no. The protocol in $[23]$ presents a quantum election scheme without complicated measurements: this scheme uses a Fourier transform for executing voting and can be implemented as soon as the implementation of the discrete Fourier transform becomes possible. ${[24]}$ proposes a new protocol for quantum anonymous voting, which protects both the voters from a curious tallyman and all the participants from a dishonest voter in an unconditional way. The voting for the candidates of the ballots is still restricted to yes or no. $[25]$ presents a traveling ballot scheme based on quantum mechanics, the main idea of this scheme is that the voters cast their votes in an orderly manner with a traveling quantum state. In this scheme, the voters can vote for many candidates, and they can determine whether to cast their ballots to the traveling state. As there is still no reliable way to avoid the voters from voting more than once, a malicious voter may try to detect the execution of the election. All the protocols mentioned above are based on entangled quantum states.

Unlike these protocols, in paper $[26]$, Okamoto et al. present a relatively expedient quantum voting scheme based on conjugate coding, in which a ballot is an unknown quantum state that enables a voter to exercise his right to vote. This scheme ensures the unconditional security and anonymity of the election without the use of entangled quantum states, and the quantum blank votes generated in advance avoid a voter from voting more than once. As the quantum ballot is randomized by the voter before sending it to the voting administrator, nobody can trace the voter's ballot. This efficiently protects the private of the voter, but at the same time makes it impossible for the voter to check whether he has voted successfully or not; that is, verifiability is not guaranteed. On the basis of this protocol, in paper $[27]$, we present a new quantum election scheme, which depends on the security of the anonymous quantum key distribution to ensure unconditional security. This scheme ensures the completeness, soundness, privacy, eligibility, unreusability, fairness and verifiability of an election while the voting administrator and the counter are semi-honest; it can efficiently avoid a voter from voting more than once and works even when there exist losses and errors in the quantum channels. However, in this scheme, the security of the election depends considerably on the credibility of the voting administrator and the counter, as the administrator may try to forge valid votes by impersonating the voters and there is no reliable way to solve the dispute between the administrator and the voters. Further, the security will be threatened by a collusion of the administrator and the counter.

A distributed scheme can be used for solving a dispute between the voting administrator and the voters. The distributed scheme is a scheme in which several independent parties, e.g., several candidates of the election, collaborate to act as the voting administrator. A combination of a traditional one-time pad and a distributed scheme can efficiently ensure information security and avoid the voting administrator from impersonating a voter. However, the security of the protocol is difficult to achieve in real life because there is no effective way to guarantee that the parties that form the voting administrator will not cooperate to trace the ballots forever. Whenever they cooperate, the privacy of the election is at risk even when there are a sufficient number of key strings. In this paper, we propose a new distributed quantum election scheme, in which we use a combination of a distributed scheme and quantum cryptography to construct an unconditionally secure distributed anonymous quantum key distributed scheme and to remove the threat posed by the voting administrator and the counter. The security of the anonymous quantum key distribution is based on the security of the quantum key distribution. With the help of the voting administrator, the voter can anonymously establish a key string with the counter; this key is invisible to the administrator. In the new distributed scheme, when the election is completed, nobody can trace the ballot to detect the privacy of the election even if the voting administrator and the counter collaborate to do so; this to an extent improves the security level of the scheme.

The rest of this paper is organized as follows: In Section 2, we present our former quantum election scheme based on anonymous quantum key distribution, and then we discuss a traditional election scheme that uses the distributed scheme and analysis its security. In Section 3, we present distributed anonymous quantum key distribution schemes that will be used in the distributed quantum election scheme proposed in the next section. We present the proposed distributed quantum election scheme in Section 4, and in Section 5, we discuss the advantages of the proposed quantum election scheme. Finally, we present our conclusions in Section 6.
\section{Preliminary}
We use the notation $\|$ to denote the concatenation of strings. $E_k[M]$ denotes an unconditionally secure symmetric encryption algorithm, and $f(\cdot)$ denotes an information-secure one-way function:
\begin{equation}
F(a_i,b_i)=a_i\oplus b_i,
\end{equation}
where $a_i$ and $b_i$ are bit strings having the same length.

In view of the properties described in $[1-3]$, a secure quantum election scheme should satisfy following: It is complete, if one ballot is valid, it should be countable. It should be sound so that a dishonest voter cannot disturb the election. It is anonymous, the owner of a ballot is invisible to others. It should be non-repeatable, and hence, no voter can vote successfully twice. It should be fair so that the earlier voters cannot affect the later voters. It should be verifiable, a voter should be able to check his ballot at the end of the election. We use ${ID}_i$ to represent the identity of the eligible voter $V_i$.

As introduced in $[28]$, a scheme with covert security can guarantee that once an adversary attempts to cheat in order to destroy some security properties of the scheme, the honest parties will notice the cheating attempt with some constant probability. In other words, any irregularity in the scheme should be detected with some constant probability. We believe that a distributed quantum election scheme in a sense should ensure covert security.
\subsection{Quantum election based on anonymous quantum key distribution}
We presented an election scheme based on an anonymous quantum key distribution scheme using a semi-honest model in $[27]$, this scheme can efficiently satisfy all the properties mentioned above. Four phases are included in the scheme: initial phase, authentication phase, key distribution phase and voting phase. Several voters $V_j$, j=1,2,$\cdots$,N, the voting administrator Bob, and the counter Charlie are also involved.

\textbf{Initial phase:}
In the initial phase, the voting administrator Bob publishes a set $\mathcal{S}\subset \{0,1\}^s$. Each element of the set is randomly chosen by Bob to represent an eligible candidate. In the election scheme, an eligible voter $V_j$ chooses an element as his ballot $v_j$.

The voting administrator Bob establishes a key $k_{bj}$ with every eligible voter $V_j$, j=1,2,$\cdots$,N, by directly contacting or using an unconditionally secure quantum key distribution protocol. All the four parts of $k_{bj}$ are selected uniquely for $V_j$.

All these tasks should be completed in advance.

\textbf{Authentication phase:}
When the eligible voter $V_j$ wants to vote, he sends a request by sending the group $(ID_j,k_j)$ to Bob. Then Bob checks whether $V_j$ has successfully applied for voting before. If not, Bob verifies whether the string $k_j$ is correct: if it is correct, Bob accepts $V_j$'s request.

At the end of the authentication phase, Bob announces the number of verified voters(we denote it by n) and publishes a set that contains all the verified $ID_j$, $j\in\{1,2,\cdots,N\}$. Now the scheme turns to the key distribution phase.

\textbf{Key distribution phase:}
In this phase, Bob helps each verified voter $V_j$ to execute an anonymous quantum key distribution protocol to establish a key $K_{ic}$ between $V_j$ and Charlie. Here $K_{ic}=K_{icL}\|K_{icR}$. The anonymous quantum key distribution scheme is unconditional secure under semi-honest model, and $V_j$ can verifies whether the anonymous quantum key distribution process is successful.

\textbf{Voting phase:}
While $V_j$ ensures that the anonymous quantum key distribution is successful, he has anonymously established a 2s-bit key $K_{ic}$ with Charlie successfully. Then, he chooses an element from set $\mathcal{S}$ as his ballot $v_j$ and uses $K_{icR}$ to encrypt his ballot. Next, he anonymously sends the encrypted ballot along with $K_{icL}$ to Charlie.

Charlie checks whether $K_{icL}$ is correct and whether he has accepted it before. If it is correct and he has not accepted it before, he extracts the corresponding $K_{icR}$ and uses it to decrypt the encrypted ballot. If the outcome $v_j\in \mathcal{S}$, Charlie counts $v_j$ and accepts $K_{icL}$.

While all the verified voters vote successfully, Charlie counts the number of each candidate's ballots. Subsequently, he randomly arranges all the accepted groups $(K_{icL},v_j)$ and publicly publishes them for the voters to trace their ballots. The scheme is now completed.

%This quantum election scheme can ensures the completeness, soundness, privacy, eligibility, unreusability, fairness and verifiability of a large-scale election, and it can works even if there exist loss and errors in quantum channels. In addition, any irregularity in this scheme is sensible. However,in this scheme we assume Bob and the Charlie are semi-honest. Actually, a malicious administrator Bob or a malicious counter Charlie may also try to forge valid votes: as Bob knows the key pre-shared between the voters and Charlie, he can easily personate the eligible voter $V_j$ and vote successfully; in addition, a malicious Charlie may tamper the ballot of an eligible voter. Although these irregularities will be discovered by $V_j$ at the end of the scheme, there is no way to distinguish between a dishonest voter and a malicious Bob or a malicious Charlie. To avoid this, we can use a distributed scheme, in which several independent parties, e.g., several candidates of the election, collaborate to act as the voting administrator and another several independent parties collaborate to act as Charlie. They won't cooperate up forging invalid votes or tracing valid ballots. The distributed scheme ensures the credibility of the voting administrator and the counter.

The quantum election scheme satisfies all the properties mentioned above efficiently, and any irregularity in the scheme is sensible while the administrator and the counter are semi-honest. When an attacker attempts to impersonate a voter to vote, he will be detected by the voter. However, in an election scheme, the administrator and the counter may also try to adversely affect the election: If Bob is malicious, he can easily impersonate eligible voters and help a candidate to forge ballots; at the same time, a malicious counter may also tamper the ballot of an eligible voter. Although eligible voters can discover these irregularities, there is no reliable way to prove their discovery. In order to solve this problem, we propose a new quantum election scheme, which combines the distributed scheme with quantum cryptography to improve the security level of the quantum election. In the distributed scheme, the voting administrator is made up of several independent parties who will not collaborate to adversely affect the election during the scheme. While the scheme is completed, the security and privacy of the voters will not be threatened by the voting administrator and the counter, even if the two parties collaborate.
\subsection{Traditional distributed election scheme}
Durette et al. use a combination of public key cryptography and multiple administrators to improve the security of the overall voting system by avoiding a single administrator from forging valid votes in $[29]$; $[30]$ presented a scheme in which the work for a voter is linear in the number of authorities but can be instantiated to yield information-theoretic privacy. When there are $n$ authorities, $m$ voters, the security parameter is $k$, the total amount of communication will be $O(kmn)$ bits, and the required effort for any authority and any voter will be $O(km)$ and $O(kn)$ operations, respectively. In this scheme for any threshold $t\leqslant n$, privacy will be assured against coalitions that include at most $t-1$ authorities, and robustness against coalitions that includes at most $n-t$ authorities. An information security traditional distributed election scheme can be described as follows:

As mentioned in the previous paper, there are three parties involved in the scheme: the voters, the voting administrator Bob, and the counter Charlie. In particular, the voting administrator Bob is made up of multiple independent entities. For the sake of simplicity, we assume that it is made up of two independent entities $Bob_1$ and $Bob_2$, who will not cooperate to cheat. Before the voting, the voting administrator publishes a set $\mathcal{Y}\subset \{0,1\}^y$. Each element of the set is a y-bit string that randomly chosen by Bob to represent an eligible candidate.

The voting administrator establishes a secret number $r_i$ to each eligible voter in advance. The secret number is visible to both $Bob_1$ and $Bob_2$.

\subsubsection{Initial phase}

(1) $V_i$ applies for voting by sending his/her identity $ID_i$ along with the secret number $r_i$ to Bob.

(2) After getting $V_i$'s request, Bob checks whether $V_i$ has applied for voting before. If he has, Bob rejects his request; otherwise, Bob checks whether $r_i$ is correct. If it is correct, the scheme moves to the next step.

(3) $Bob_1$ and $Bob_2$ respectively establish a secret string with $V_i$ by directly contacting or using an unconditionally secure quantum key distribution protocol. We denote these strings by
\begin{gather}
S_{i1}=N_{i1}\|T_{i1},\\
S_{i2}=N_{i2}\|T_{i2},
\end{gather}
and $V_i$ uses the function $f(\cdot)$ to generate
\begin{gather}
N_i=f(N_{i1},N_{i2}),\\
T_i=f(T_{i1},T_{i2}).
\end{gather}

(4) $Bob_1$ and $Bob_2$ respectively establish the secret strings $S_{i1}$ and $S_{i2}$ with Charlie in the same manner as that used in the previous step, and Charlie also uses function $f(\cdot)$ to get the strings $N_i$ and $T_i$.

\subsubsection{Voting and counting phase}

(1) $V_i$ chooses one candidate as his vote $v_i$ and encrypts it by $N_i$. Then, he sends $E_{N_i}[v_i]$ to Charlie along with $T_i$.

(2) When he receives $(T_i, E_{N_i}[v_i])$, Charlie checks whether he has received $T_i$ before. If not, he uses $T_i$ to get the corresponding $N_i$ recorded in his database and decrypts $E_{N_i}[v_i]$ to get the plaintext $v_i$. After extracting $v_i$, Charlie checks whether $v_i$ is correct. If it is correct, Charlie counts this vote.

After all the votes have been counted, Charlie publishes all the groups $(T_i,v_i)$ for the eligible voter to check whether he/she has voted successfully.

The scheme is now completed.

\subsubsection{Security analysis}

In this election scheme, the voting administrator will check the identity of each voter before accepting a voting request and will not respond to one voter twice; Charlie will check whether he has received $T_i$ before counting $v_i$ so that the eligibility and unreusability criteria are satisfied. At the same time, the earlier voters¡¯ votes will have no effect on the later ones, and all the eligible votes will be count correctly. Hence, the fairness and completeness criteria are satisfied. Charlie will check each ballot before counting it, thus guaranteeing the soundness of the process. At the end of the scheme, Charlie will publish all the counted ballots for each voter to check whether he/she has voted successfully; hence, the scheme satisfies the verifiability criterion. The properties mentioned above are easy to prove; hence, now, we mainly discuss the privacy of the scheme.

In this scheme, two independent parties collaborate to act as the voting administrator, the function $f(\cdot)$ is information security, and the final
key string $N_i$ and $T_i$ between $V_i$ and Charlie is invisible to others(the independent parties $Bob_1$ and $Bob_2$ are included). As
long as one of the two parties does not cooperate with the other, the communication key strings $N_i, T_i$ are secure. However, the security of
the protocol is difficult to achieve in real life because there is no effective way to guarantee that the two parties $Bob_1$ and $Bob_2$ will
 not cooperate to trace the ballots forever. Whenever the two parties cooperate, the privacy of the election is compromised.
%An ideal election protocol should be information security and guarantee that the communication key strings are invisible to others at any time.

\section{Distributed anonymous quantum key distribution}

Considering the problem that mentioned in the case of the traditional distributed election scheme, it is impossible to ensure that the two distributed parties $Bob_1$ and $Bob_2$ will not cooperate forever. However, according to common sense, in a real election, we can think that there exists an overseeing body to monitor the elections; this institution will supervise both sides for a certain period of time so as to ensure that they do not cooperate with one another within this time period. On the basis of this viewpoint, we assume that the administrator is made up of two parties $Bob_1$ and $Bob_2$, who cannot collaborate to cheat during the scheme. When the scheme is complete, the private of the voter will not be compromised even if $Bob_1$ and $Bob_2$ cooperate.

Anonymous quantum key distribution(AQKD) can be used for ensuring the private of the voter in the case of a semi-honest model$^{[27]}$. If one voter can share a key string with Charlie anonymously, he/she can easily encrypts his/her ballot anonymously. Once others cannot match the key string with the voter, they also cannot trace the ballot. In view of the problems in the earlier paper$^{[27]}$, we present an improved distributed AQKD protocol that will be used in the new distributed quantum election scheme.
\subsection{Qubit-based distributed AQKD protocol}
Suppose a voter $V_i$ wants to anonymously establish a key string with the counter Charlie with the help of the voting administrator Bob(made up of two independent parties $Bob_1$ and $Bob_2$). The voter wants to ensure that nobody except himself/herself and Charlie can get the key string, which is difficult to achieve through traditional key distribution, because there is no reliable way to avoid others from copying the information. The quantum non-cloning theorem guarantees that it is impossible to measure or copy an unknown qubit without being detected. Based on this, we present a qubit-based AQKD protocol:

Prerequisite: $Bob_1$ has established a secret string $S_{i1}=(N_{i1}\|T_{i1})$ with $V_i$ and Charlie, and $Bob_2$ has established a secret string $S_{i2}=(N_{i2}\|T_{i2})$ with $V_i$ and Charlie. Both $V_i$ and Charlie compute $N_i=f(N_{i1},N_{i2})$, $T_i=f(T_{i1},T_{i2})$.

\textbf{Step 1.} $V_i$ randomly chooses a string $R_i\in\{0,1\}^m$, and generates his qubits $|\alpha_i\rangle$ by the method of conjugate coding$^{[31]}$:
\begin{equation}
|\alpha_i\rangle=H^{N_i}|R_i\rangle=\otimes_{j=1}^m (H^{N_i^j}|R_i^j\rangle),
\end{equation}
where $N_i^j$ and $R_i^j$ denote the $j$-th bit of $N_i$,$R_i$, and $H^{N_i}=\otimes_{j=1}^m H^{N_i^j}$,
$H^0=I=\left(
                   \begin{array}{cc}
                     1 & 0 \\
                     0 & 1 \\
                   \end{array}
                 \right)
$, $H^1=H=\dfrac{1}{\sqrt{2}}\left(
                               \begin{array}{cc}
                                 1 & 1 \\
                                 1 & -1 \\
                               \end{array}
                             \right)
$.

Then, $V_i$ anonymously transmits the quantum state $|\alpha_i\rangle$ with the secret string $T_i$ to Charlie.

\textbf{Step 2.} Charlie checks whether he has received $T_i$ before. If not, he uses $T_i$ to get the corresponding string $N_i$, and measures the quantum state $|\alpha_i\rangle$ depending on the value of $N_i$: if $N_i^j=0$, he measures the qubit $|\alpha_i^j\rangle$ with the rectilinear basis $\{|0\rangle,|1\rangle\}$; otherwise he measures it with the diagonal basis $\{|+\rangle,|-\rangle\}$. After obtaining  the outcome, Charlie publishes a subset of the outcome with $T_i$ and all the location information of the bits of the subset; we denote the subset by $\sigma_i$.

\textbf{Step 3.} $V_i$ checks whether the subset $\sigma_i$ is equal to the corresponding subset of $R_i$. If it is, $V_i$ knows that the quantum key distribution is successful. Then, he/she deletes the checking bits $\sigma_i$ and extracts the remaining bits of $R_i$ as the final encrypted key string
\begin{equation}
K_i=R_i-\sigma_i=K_{iL}\|K_{iR},
\end{equation}
now $V_i$ has anonymously established a key string $K_i$ with Charlie.

\emph{\textbf{Security analysis}}

In this protocol, we assume that $Bob_1$ and $Bob_2$ will not cooperate until the protocol is complete, and the function $f(\cdot)$ is information security; thus, neither $Bob_1$ nor $Bob_2$ can get the strings $N_i$ or $T_i$ during the AQKD protocol. The quantum state $|\alpha_i\rangle$ is an unknown quantum state to others($Bob_1,Bob_2$ included); hence, it is impossible to make a copy of the strings. A malicious entity has to measure the initial qubits if he wants to get some useful information about the final encryption key string. However, without the encoding basis of the qubits, he will choose the wrong measuring basis with a $50\%$ probability for each qubit and then introduce no less than $25\%$ error rate in Charlie's measuring outcome. This implies that without $T_i$ it is impossible to make a correct measurement on the quantum state $|\alpha_i\rangle$, and an incorrect measurement will affect Charlie's measuring outcome and be detected by $V_i$ in the last step of the anonymous key distribution phase; $V_i$ will then stop the scheme. As a result, nobody except $V_i$ and Charlie can get the final encryption key string $K_i$. The probability that others except $V_i$ and Charlie guess the correct x-bit key string $K_i$ is not more than $(\frac{1}{2})^x$. When x is sufficiently large, the probability is negligible. In other words, when the anonymous key distribution phase is complete, there is no effective way for $Bob_1$ and $Bob_2$ to get the final encryption key string even if they cooperate to do so.

\emph{\textbf{Practicability analysis}}

Considering the problem mentioned in the case of the traditional distributed election scheme, we just assume that the two independent entities $Bob_1$ and $Bob_2$ will not cooperate for a limited period of time(just limited to the end of the election). This is reasonable and easy to achieve in practice.

This AQKD protocol can also work even if there exist losses and errors in the quantum channels. In order to establish an error-free key, Charlie can publish all the qubits that he has received and then publish the check string and its serial numbers in Step 2 of the anonymous key distribution phase. If the error rate of the check string is acceptable, $V_i$ can ensure that the key distribution is successful. After verifying that the key distribution is successful, $V_i$ extracts the string $K_i$ from $R_i$ and subsequently uses $K_i$ and a key redistribution protocol$^{[32]}$ for establishing the final error-free key and then uses the error-free key to encrypt his/her ballot$^{[27]}$.

\subsection{Qubit-string-based distributed AQKD protocol}
The former qubit-based distributed AQKD protocol can efficiently avoid Bob from impersonating the voters to vote or tracing the ballots after the election. The combination of anonymous quantum key distribution and the distributed scheme guarantees the private of the voter. When the protocol is completed, neither $Bob_1$ nor $Bob_2$ can get the final encrypted key string even if they collaborate to do so. However, there is a further problem: in this scheme Bob and Charlie should be two independent parties that will never cooperate. Once the two parties cooperate the privacy of the voter is not guaranteed. An ideal AQKD protocol should ensure that once the protocol is completed nobody can compromise the private of the voter even if the voting administrator and the counter cooperate to do so.

A qubit-string-based scheme can be used for solving this problem. The improved AQKD protocol is as follows:

Prerequisite: $Bob_1$ and $Bob_2$ respectively establish specific key strings with $V_i$, we denote the key strings by $N^{(1)},L^{(1)}$ and $N^{(2)},L^{(2)}$; at the same time, $Bob_1$ and $Bob_2$ respectively establish a specific key string with Charlie, we denote the two strings by $C^{(1)}$ and $C^{(2)}$. Here,
\begin{equation}
C^{(1)}=S^{(1)}\|M^{(1)}\|N^{(1)}\|L^{(1)},\\\\C^{(2)}=S^{(2)}\|M^{(2)}\|N^{(2)}\|L^{(2)},
\end{equation}
where
\begin{gather}
N^{(1)},N^{(2)}\in\{0,1\}^m,M^{(1)},M^{(2)}\in\{0,1\}^l,L^{(1)},L^{(2)}\in\{0,1\}^l,\\
S^{(1)}=(a_{11},\cdots,a_{1m},\cdots,a_{l1},\cdots,a_{lm})\in\{0,1\}^{lm},\\
S^{(2)}=(b_{11},\cdots,b_{1m},\cdots,b_{l1},\cdots,b_{lm})\in\{0,1\}^{lm}.
\end{gather}
The strings are the same for every voter.

\textbf{Step 1.} $Bob_1$ and $Bob_2$ respectively use the random number generator to prepare a random string
\begin{align}
R_i^{(1)}=(c_{11},\cdots,c_{1m},\cdots,c_{l1},\cdots,c_{lm})\in\{0,1\}^{lm},\\
R_i^{(2)}=(d_{11},\cdots,d_{1m},\cdots,d_{l1},\cdots,d_{lm})\in\{0,1\}^{lm},
\end{align}
where
\begin{align}
c_{j1}\oplus c_{j2}\oplus \cdots \oplus c_{jm}=M^{(1)}_j,0<j<l+1,\\
%d_{j1}\oplus d_i^2\oplus \cdots \oplus d_i^m=M_2^i,0<i<l+1,
d_{j1}\oplus d_{j2}\oplus \cdots \oplus d_{jm}=M^{(2)}_j,0<j<l+1,
\end{align}
and collaborate to generate a quantum state $|\alpha_i\rangle$ by using conjugate coding:
\begin{equation}
\begin{split}
|\alpha_i\rangle&=H^{S^{(2)}}Y^{R^{(2)}}H^{S^{(1)}}Y^{R^{(1)}}|0\rangle\\
                &=\otimes _{j=1}^{l} |\alpha_{ij}\rangle\\
                %&=\otimes _{j=1}^{l} H_{Bob_2}^{b_j}Y_{Bob_2}^{d_j}H_{Bob_1}^{a_j}|c_j\rangle\\
                &=\otimes _{j=1}^{l}\otimes _{k=1}^mH^{b_{jk}}Y^{d_{jk}}H^{a_{jk}}|c_{jk}\rangle.
\end{split}
\end{equation}
Here $|0\rangle$ denotes an $lm$-dimensional zero vector,
and $H^0=I=\left(
                   \begin{array}{cc}
                     1 & 0 \\
                     0 & 1 \\
                   \end{array}
                 \right)
$, $H^1=H=\dfrac{1}{\sqrt{2}}\left(
                               \begin{array}{cc}
                                 1 & 1 \\
                                 1 & -1 \\
                               \end{array}
                             \right)
$,$Y^0=I=\left(
                   \begin{array}{cc}
                     1 & 0 \\
                     0 & 1 \\
                   \end{array}
                 \right)
$, $Y^1=Y=\left(
                               \begin{array}{cc}
                                 0 & -1 \\
                                 1 & 0 \\
                               \end{array}
                             \right)
$.

Then Bob sends the quantum state to $V_i$ via a secure quantum channel. In order to avoid the attacker from intercepting the qubits to extract information, we can assume that both $Bob_1$ and $Bob_2$ use a secret key to flip the qubits and the voter uses the keys to get the initial qubits.

\textbf{Step 2.} After receiving $|\alpha_i\rangle$, $V_i$ randomly chooses a strings $K_i\in\{0,1\}^{l-m}$ and computes
\begin{equation}
N=f(N^{(1)},N^{(2)}),L=f(L^{(1)},L^{(2)}), T_i=E_L[K_i\|N],
\end{equation}
then, he randomly chooses a string $P_i\in\{0,1\}^{lm}$, where
\begin{equation}
P_i=(e_1,e_2,\cdots,e_l)=(e_{11},\cdots,e_{1m},\cdots,e_{l1},\cdots,e_{lm}),
\end{equation}
and
\begin{equation}
e_{j1}\oplus e_{j2} \oplus \cdots \oplus e_{jm}=(T_i)_j, \text {for} \   0<j<l+1.
\end{equation}

\textbf{Step 3.} $V_i$ generates
\begin{equation}
%\begin{split}
|\alpha_i\rangle^{\prime}=Y^{P_i}|\alpha_i\rangle=\otimes _{j=1}^{l}\otimes _{k=1}^mY^{e_{jk}}|\alpha_i\rangle_{jk}.
%\end{split}
\end{equation}
Then $V_i$ sends $|\alpha_i\rangle^{\prime\prime}$ to Charlie anonymously via a secure quantum channel.% along with a random string $Q_i$ which is transformed to qubits according to the rectilinear basis $\{|0\rangle,|1\rangle\}$.

\textbf{Step 4.} Charlie computes $M=f(M^{(1)},M^{(2)})$, $N=f(N^{(1)},N^{(2)})$, $s=f(S^{(1)},S^{(2)})$ and $L=f(L^{(1)},L^{(2)})$ in advance. When he receives $|\alpha_i\rangle^{\prime}$, Charlie measures it depending on the value of the string $s$: if $s_j=0$, he measures the qubit $|\alpha_{ij}\rangle^{\prime}$ with the rectilinear basis $\{|0\rangle,|1\rangle\}$, where j$\in\{1,2,\cdots,lm\}$ throughout; otherwise he measures it with the diagonal basis $\{|+\rangle,|-\rangle\}$. After getting the outcome $r^\prime$, where
\begin{equation}
r^\prime=(h_{11},h_{12},\cdots,h_{1m},\cdots,h_{l1},\cdots,h_{lm}).
\end{equation}
Charlie computes $M^\prime=(M_1^\prime,M_2^\prime,\cdots,M_l^\prime)$ with
\begin{equation}
M_j^\prime=r_{j1}^\prime \oplus \cdots \oplus r_{jm}^\prime, j=1,2,\cdots,l
\end{equation}
and uses the string M to extract
\begin{equation}
T_i^\prime=M\oplus M^{\prime},
\end{equation}
then, he can obtain the strings $K_i^{\prime}\|N^{\prime}$ by decrypting $T_i^{\prime}$ with the key $L$. Then he checks whether $N^{\prime}$ is correct. If it is correct, Charlie ensures that he gets the correct key string $K_i$.

After verifying all the key strings, Charlie will publicly publish a subset of the string for the voter to verify whether the anonymous quantum key distribution is successful; every voter publishes whether the quantum key distribution is successful, and the voting administrator helps the failed ones to restart a new anonymous quantum key distribution.

\emph{\textbf{Security and practicability analysis}}

In this protocol we use m-qubits to transmit one key bit. Further, the quantum state $|\alpha_i\rangle^{\prime\prime}$ has been randomized by $V_i$, and hence, the voting administrator and the counter cannot match the key string $K_i$ with the voter $V_i$ even if they collaborate to do so.

When there exist losses and errors in the quantum channels, we can improve the protocol as follows:

In the voting phase, $V_i$ uses the random number generator to generate a random number
\begin{equation}
P_i=(e_1,e_2,\cdots,e_l)=(e_{11},\cdots,e_{1m},\cdots,e_{l1},\cdots,e_{lm}),
\end{equation}
then he uses a classic error correction coding(ECC) to encode it and get the corresponding code $D_i$.

When he receives the qubits $|\alpha_i\rangle$ from Bob, $V_i$ adds an operation as follows:
\begin{equation}
|\alpha_i\rangle^{\prime}=\otimes_j Y^{D_{ij}} |\alpha_{ij}\rangle.
\end{equation}
$V_i$ adds $Y^{D_{i*}}$ to $|\alpha_{i*}\rangle$ only if he receives this qubit. After doing so, $V_i$ generates the key string $p_i$, where \begin{equation}
p_{ij}=e_{j1}\oplus e_{j2}\oplus \cdots \oplus e_{jm}, j=1,2,\cdots,l,
\end{equation}
then he sends $|\alpha_i\rangle^{\prime}$ to Charlie along with the serial numbers of the qubits that he receives from Bob. Charlie measures the qubits and extracts the code $D_i^\prime=D_i\oplus e$. He uses $D_i^\prime$ to recover the string $P_i$ and then generates the key string $p_i$. A dishonest voter or an attacker may also try to forge a ballot. As he/she is not aware of the string $S$ and $M$, he/she will introduce no less than 75\% error rate in the string $D_i^\prime$. While the error correction ability of the ECC is not available to correct such a large error rate, it is impossible to recover the string $P_i$.

\section{Quantum distributed election schemes}
A complete voting process in our quantum election scheme includes four phases: initial phase, anonymous quantum key distribution phase, voting phase and counting phase. Several voters $V_j$, j=1,2,$\cdots$,N, the voting administrator Bob(made up of two independent parties $Bob_1$ and $Bob_2$), and the counter Charlie are also involved.

If Bob and Charlie are two independent parties that will not collaborate forever, we can use both the qubit-based distributed AQKD protocol and the qubit-string-based AQKD protocol that presented in Section 2; in order to ensure that the private of the voter will not be threatened by the collaboration of the voting administrator Bob and the counter Charlie forever, we use the qubit-string-based AQKD protocol in our distributed quantum election scheme.

\subsection{Initial phase}
$Bob_1$ and $Bob_2$ respectively establish a specific key string with every eligible voter, we denote the key strings by $N^{(1)},L^{(1)}$,$N^{(2)},L^{(2)}$. At the same time, $Bob_1$ and $Bob_2$ respectively establish a specific key string with Charlie, we denote these two strings by $C^{(1)}$,$C^{(2)}$. Here
\begin{equation}
C^{(1)}=S^{(1)}\|M^{(1)}\|N^{(1)}\|L^{(1)},\\\\C^{(2)}=S^{(2)}\|M^{(2)}\|N^{(2)}\|L^{(2)},
\end{equation}
where $N^{(1)},N^{(2)}\in\{0,1\}^m, M^{(1)},L^{(1)},M^{(2)},L^{(2)}\in\{0,1\}^l$.

\subsection{Anonymous key distribution phase}
In this phase each eligible voter $V_i$ anonymously establishes an encrypted key string $K_i=K_{iL}\|K_{iR}$ with the counter Charlie by using the qubit-string-based AQKD protocol. The key string is used for encrypting the ballot and it is invisible to the others($Bob_1$ and $Bob_2$ included) except for $V_i$ and Charlie.

After all the voters anonymously establish key strings with Charlie, the scheme moves to the voting phase.

\subsection{Voting and counting phase}

(1) $V_i$ chooses a candidate as his vote $v_i$ and encrypts it with $K_{iR}$. Then, he sends $E_{K_{iR}}[v_i]$ to Charlie along with $K_{iL}$.

(2) After receiving $(K_{iL},E_{K_{iR}}[v_i])$, Charlie checks whether he has received $E_{K_{iL}}$ before. If the check succeeds, Charlie uses $K_{iL}$ to extract the corresponding $K_{iR}$. Then, he decrypts $E_{K_{iR}}[v_i]$ to get the ballot $v_i$. If the outcome $v_i$ is eligible, Charlie counts this vote.

After all the votes has been count, Charlie publishes all the groups ${(E_{K_{iL}},v_i)}$ for the eligible voter to check whether he/she has voted successfully.

The scheme is now completed.

\subsection{Security and practicability analysis}
\textbf{Privacy:} The quantum state $|\alpha_i\rangle^{\prime}$ is randomized by $V_i$ before it is sent to Charlie. The initial quantum state $|\alpha_i\rangle$ is generated by both $Bob_1$ and $Bob_2$. After $Bob_1$'s operation, the qubits can be denoted as
\begin{equation}
|\alpha_i^{(1)}\rangle=H^{S^{(1)}}Y^{R_i^{(1)}}|0\rangle,
\end{equation}
then the qubits are transmitted to $Bob_2$. As the strings $S^{(1)},R_i^{(1)}$ are not aware to $Bob_2$, the density matrix of the qubits can be expanded as
\begin{equation}
%\begin{split}
\rho_i^{(1)}=\frac{1}{2^{2lm}}\sum_{S^{(1)},R_i^{(1)}}H^{S^{(1)}}Y^{R_i^{(1)}}\sigma Y^{R_i^{(1)}}H^{S^{(1)}},
%\end{split}
\end{equation}
where $\sigma$ denotes the density matrix of $|0\rangle$.
$Bob_2$ flips the quantum state depending on the value of $S^{(2)},R_i^{(2)}$, as $S^{(2)},R_i^{(2)}$ are randomly distributed, after $Bob_2$'s operation, the density matrix of the qubits $|\alpha_i\rangle$ that transmitted to $V_i$ can be written as
\begin{equation}
\begin{split}
\rho_i&=\frac{1}{2^{4lm}}\sum_{S^{(1)},S^{(2)},R_i^{(1)},R_i^{(2)}}H^{S^{(2)}}Y^{R_i^{(2)}}H^{S^{(1)}}Y^{R_i^{(1)}}\sigma Y^{R_i^{(1)}}H^{S^{(1)}}Y^{R_i^{(2)}}H^{S^{(2)}}\\
    &=\frac{1}{2^{4lm}}\sum_{S^{(1)},S^{(2)},R_i^{(1)},R_i^{(2)}}H^{S^{(1)}\oplus S^{(2)}}Y^{R_i^{(1)}\oplus R_i^{(2)}}\sigma Y^{R_i^{(1)}\oplus R_i^{(2)}}H^{S^{(1)}\oplus S^{(2)}}\\
    &=\frac{1}{2^{2lm}}\sum_{s,r_i}H^{s}Y^{r_i}\sigma Y^{r_i}H^s,
\end{split}
\end{equation}
here $s=f(S^{(1)},S^{(2)}), r_i=f(R_i^{(1)},R_i^{(2)})$.
We can verify that the set of $2^{2lm}$ unitary matrices $\{H^sY^{r_i}\}$ forms an orthonormal  basis. In view of the quantum one-time pad in $[11,12]$, in this basis the density matrix $\sigma$ can be expanded as
\begin{equation}
\sigma=\sum_{\eta,\xi}a_{\eta,\xi}H^{\eta}Y^{\xi},
\end{equation}
where $a_{\eta,\xi}=Tr(\sigma Y^{\xi}H^{\eta})/2^{lm}$. Now the density matrix $\rho$ can be expanded as
\begin{equation}
\begin{split}
\rho_i&=\frac{1}{2^{2lm}}\sum_{s,r_i}H^sY^{r_i}\sum_{\eta,\xi}a_{\eta,\xi}H^{\eta}Y^{\xi}Y^{r_i}H^s\\
    &=\frac{1}{2^{2lm}}\sum_{\eta,\xi}a_{\eta,\xi}\sum_{s,r_i}H^sY^{r_i}H^{\eta}Y^{\xi}Y^{r_i}H^s\\
    &=\frac{1}{2^{2lm}}\sum_{\eta,\xi}a_{\eta,\xi}\sum_{s,r_i}(-1)^{r_i\cdot\eta\oplus s\cdot\xi}H^{\eta}Y^{\xi}\\
    &=\sum_{\eta,\xi}a_{\eta,\xi}\delta_{\eta,0}\delta_{\xi,0}H^{\eta}Y^{\xi}\\
    &=a_{0,0}I=\frac{Tr(\sigma)}{2^{lm}}=\frac{1}{2^{lm}}I.
\end{split}
\end{equation}
After receiving the quantum state, $V_i$ adds quantum operation to it depending on the value of the strings $P_i$. We use $\rho_i^{\prime}$ to denote the density matrix of the randomized quantum state. As $P_i$ is randomly distributed, after $V_i$'s operation, for the attacker who cannot obtain the generation basis and values, the density matrix of the qubits can be expanded as
\begin{equation}
\rho_i^{\prime(0)}=\frac{1}{2^{2lm}}\sum_{s,r^{(0)}}H^sY^{r^{(0)}}\sigma Y^{r^{(0)}}H^s
             =\frac{1}{2^{lm}}I,
\end{equation}
where $r_i^{(0)}=r_i\oplus P_i$. This implies the information that determines the unitary transformation $Y$ after $V_i$'s operation. For $Bob_1$ and $Bob_2$, the density matrix of the qubits can be respectively expanded as
\begin{gather}
\rho_i^{\prime(1)}=\frac{1}{2^{2lm}}\sum_{S^{(2)},r^{(1)}}H^{S^{(2)}}Y^{r^{(1)}}\sigma Y^{r^{(1)}}H^{S^{(2)}}=\frac{1}{2^{lm}}I\  \text{for}\  Bob_1, \\
\rho_i^{\prime(2)}=\frac{1}{2^{2lm}}\sum_{S^{(1)},r^{(2)}}H^{S^{(1)}}Y^{r^{(2)}}\sigma Y^{r^{(2)}}H^{S^{(1)}}=\frac{1}{2^{lm}}I\  \text{for}\  Bob_2,
\end{gather}
where $r^{(1)}=r^{(0)}\oplus R_i^{(1)}$, and $r^{(2)}=r^{(0)}\oplus R_i^{(2)}$. The two density matrices are in a completely mixed state. Hence, $Bob_1$ and $Bob_2$ cannot obtain any useful information about $V_i's$ identity from the state. For the counter Charlie, the density matrix can be expanded as
\begin{equation}
\rho_i^{\prime(3)}=\frac{1}{2^{l(m-1)}}\sum_{r^{(3)}}Y^{r^{(3)}}\sigma Y^{r^{(3)}}=\frac{1}{2^{l(m-1)}}\sum_{r^{(3)}}|r^{(3)}\rangle\langle r^{(3)}|=\frac{1}{2^{l(m-1)}}I,
\end{equation}
where $r^{(3)}=r^{(0)}\oplus R_i^{(1)} \oplus R_i^{(2)}$=$(r^{(3)}_{11},\cdots,r^{(3)}_{1m},\cdots,r^{(3)}_{l1},\cdots,r^{(3)}_{lm})$ and satisfies the condition that
\begin{equation}
r^{(3)}_{j1}\oplus r^{(3)}_{j2}\oplus \cdots \oplus r^{(3)}_{jm}=M_j \oplus {(T_i)}_j,\  \text{for} \ 0<j<l+1.
\end{equation}

We can see that the density matrix of the final quantum state $|\alpha_i\rangle^{\prime}$ that $V_i$ sends to Charlie is a totally mixed state. Hence, the attacker(Charlie and $Bob_1, Bob_2$ included) cannot extract any useful information about $V_i$'s identity $ID_i$ even if he intercepts the entire state. At the same time, his attack will change the initial qubits and then be discovered by the honest parties. Although this discovery is not verifiable and the honest parties cannot point out the specific attacker, the attacker cannot compromise the private of a voter. Further, Charlie's measurement outcome is randomized and will not reveal any information of the voter's identity, Charlie cannot match the ballot $v_i$ with the voter $V_i$ even if he collaborates with $Bob_1$ and $Bob_2$, hence, the voters' private is guaranteed.

\textbf{Unreusability and soundness:} In the case of any two voters $V_i$ and $V_j$, Bob respectively sends a quantum state $|\alpha_i\rangle$ and $|\alpha_j\rangle$ to them. The density matrix of these two quantum states can be expressed as
\begin{gather}
\rho_j=\frac{1}{2^{2lm}}\sum_{s,r_i}H^{s}Y^{r_i}\sigma Y^{r_i}H^s=\frac{1}{2^{lm}}I,\\
\rho_j=\frac{1}{2^{2lm}}\sum_{s,r_j}H^{s}Y^{r_j}\sigma Y^{r_j}H^s=\frac{1}{2^{lm}}I,
\end{gather}
both the states are totally mixed state. Hence, the voters cannot distinguish between $|\alpha_i\rangle$ and $|\alpha_j\rangle$. Thus the voters cannot obtain any information about the preparation basis and values $R_i^{(1)}$,$R_i^{(2)}$,$S^{(1)}$ and $S^{(2)}$ of the quantum states that generated by Bob.

According to the quantum no-cloning theorem it is impossible to copy an unknown quantum state without the preparation basis of the state. If an attacker wants to forge a valid ballot, he has to prepare a quantum state and send it to Charlie. As the quantum state is randomly prepared, Charlie will get the correct value of each bit of $N^\prime$ with the probability $50\%$. Thus, the probability that a forged quantum state passes Charlie's identity check is not more than $(\frac{1}{2})^x$, while $x$ is large enough, the probability is close to 0. Hence, an attacker has to obtain the information of the state preparation basis, that is, he has to obtain the value of the string $s$.

Suppose several voters collaborate to guess the string $s$. For example, we consider the first $m$ bits of $s$. Suppose $n$ voters collaborate to guess these $m$ bits. A subset(e.g., the number is $m$) of the voter guess a random $m$-bit string and use it to measure the first m-qubit of their quantum state that received from Bob. After obtaining the outcome every voter computes the XOR of all the m bits of his/her outcome. If all the m voters obtain the same value, they believe that the string that they choose is correct; otherwise, they think that the string is wrong. The probability that the $m$ voters obtain the same XOR value is
\begin{equation}
P_1=\frac{1}{2^{m-1}},
\end{equation}
hence, they can exclude a wrong string with the probability
\begin{equation}
P_2=1-P_1=1-\frac{1}{2^{m-1}}.
\end{equation}
When $m$ is sufficiently large, the probability is close to 1. After excluding an invalid string, another set of $m$ voters can exclude another wrong string with probability $P_2$ in the same manner. In order to find the correct $s$, the voters have to exclude all the wrong strings. This requires at least $(2^m-1)*m$ voters to collaborate. When the number of the voters is considerably less than this number, it is impossible to obtain the correct generation basis even if all the voters collaborate up to find it. Without the basis the voters cannot forge quantum states to execute malicious anonymous quantum key distribution for cheating. Therefore, it is impossible for the voters to disturb an election by forging valid ballot to vote more than twice; their forged ballots will be discovered and discarded by Charlie. Hence, Soundness and unreusability are guaranteed.

\textbf{Eligibility:} Charlie will check the identity of the voter before counting his ballot; therefore, only eligible voters will be permitted to vote. At the same time, without the quantum state generated by the voting administrator, it is impossible for an attacker to forge a valid ballot; his invalid ballot will be discovered and discarded by Charlie, and hence, eligibility is guaranteed.

%after receiving $|\alpha_i\rangle$, $V_i$ executes operation to the quantum state based on the value of the two strings $P_i,T_i$ and the location of the qubit that he chosen to add his information in the subset $|\alpha_{ij}$; the string $P_i$ is randomly chosen by $V_i$, and the string $T_i$ is randomly distributed, in addition, the qubit of the subset $|\alpha_{ij}(j=1,2,\cdots,l)$ that he chooses to add his information is randomly chosen by $V_i$,

\textbf{Verifiability and fairness:} In this scheme, Charlie will publish the result of the election, so that every voter can check his/her ballot at the end of the scheme; the earlier voters have no effect on the later voters, and the scheme is fair to all the voters. Hence, verifiability and fairness are guaranteed.

\textbf{Completeness:} As the voting administrator and the counter are monitored during the election, and any dishonest attempt by the two will be detected. Further, if Charlie attempts to tamper with the statistics of the ballots, he will be detected by the voters. Hence  completeness is guaranteed.

As the qubit-string-based AQKD protocol can resist the losses and errors of the quantum channels, the distributed quantum election scheme works well even when the quantum channels have losses and errors. The only request in this scheme is the existence of the overseeing body that monitors the participants of the election during the protocol; under such a body, neither the voting administrator nor the counter can cheat during the election, thereby ensuring the security of the scheme.

\section{Discussion}

Considering the problem mentioned in the previous papers, we present a new type of unconditionally secure distributed election scheme, in which the voting administrator is made up of independent entities that cannot cooperate to cheat during an election. The security of the distributed election scheme is based on the security of the distributed anonymous quantum key distribution protocol, which depends on the quantum key distribution to ensure unconditional security. Once a voter anonymously establishes a key string with the counter, nobody can match him/her with his/her ballot as the ballot is encrypted by the key string and it is impossible to trace the key string in the anonymous quantum key distribution scheme. This distribution scheme is used for removing the threat posed by the collaboration of the voting administrator and the counter.

Compared with the traditional distributed election scheme mentioned in Section 2,  the proposed scheme can not only efficiently solve a dispute between a voter and the administrator but also solve the difficulty of monitoring the independent parties forever. In the new scheme, we just assume that there exists an overseeing body to monitor the two entities during the election, thereby guaranteeing the security and the privacy of the election. When the scheme is completed, nobody can match the key string and the corresponding voter even if the administrator and the voter collaborate to do so; hence, it is impossible to track the ballot. In the qubit-sting-based anonymous quantum key distribution scheme, we use an optical encryption method to randomize the initial quantum state and conceal the voter's identity information. After $V_i$'s operation nobody(Bob and Charlie included) can obtain any information of $V_i$. Although this anonymous quantum key distribution consumes a considerable number of key bits pre-shared between the parties, it can efficiently solve the problem that is not solved in the traditional distributed election scheme, and avoid the voting administrator and the counter from matching a voter with his ballot after the election even if they collaborate.

In view of the problems mentioned in the existing quantum election schemes$^{[21-26]}$, the proposed distributed quantum election scheme uses qubit-string-based anonymous quantum key distribution to ensure unconditional security and privacy. It ensures the completeness, soundness, privacy, eligibility, unreusability, fairness, and verifiability of the election, and removes the threat posed by the collaboration up of the administrator and the counter. Nobody can trace the ballot to destroy the privacy of the voters, and the voters can at the same time check whether they have voted successfully or not. Further, the proposed distributed quantum election scheme is relatively easy to implement because we have not used a quantum entanglement state and the scheme does not require any complicated quantum measurement.

Considering the definition of covert security discussed in $^{[28]}$, we can state that a protocol with covert security can guarantee that an honest party will notice the cheating attempt of the adversary with constant probability. In the proposed distributed quantum election scheme, the security of the voting depends on the security of the quantum key distribution; as the voting administrator
is made up of some entities that will not cooperate to cheat during the voting, a cheating attempt to forge ballots by any member of the voting administrator will be discovered by the counter in the counting phase. At the same time, a dishonest attempt of the counter to tamper with the statistics will be discovered or detected in the final counting phase. Therefore, neither the voting administrator nor the counter can cheat successfully without being discovered. Further, an attacker¡¯s attempt to forge a valid ballot will be discovered by Charlie with the probability $1-(\frac{1}{2})^x$; if $x$ is sufficiently large, the probability is close to 1. Therefore, the scheme in a sense satisfies the requirement of the security and the privacy of the information.

A more secure scheme which ensures covert security with public verification requires that the honest parties can not only discover the existence of cheat but also determine the dishonest parties or attackers. Anonymous quantum communication$^{[33,34]}$ designed for the anonymous transmission of a quantum state against an active adversary with information-theoretical security may probably be helpful to achieve this propose. Though our scheme is constructed without the use of entangle states to guarantee the practicality of the quantum election, we would like to try employing the anonymous quantum communication to construct new quantum election schemes with public verification security.

%Compared with the former election schemes, this new kind of election scheme can achieve unconditional security and privacy at the same time.

%Even if the two parties collaborate up also will be in vain. This scheme is easier to achieve in practice.

\section{Conclusion}
In this paper, we use a combination of a distributed scheme and quantum cryptography to construct an unconditionally secure distributed anonymous quantum key distributed scheme, based on which we developed a new type of unconditionally secure election scheme. In the anonymous quantum key distribution, we used the optimal encryption of quantum bits to help a voter to anonymously establish a key string with the counter and efficiently removed the threat posed by the voting administrator and the counter. In the proposed distributed quantum election scheme, after the election, nobody could trace the ballot to compromise the privacy of the voter even if the voting administrator and the counter collaborated to do so. The new distributed quantum election scheme satisfies the completeness, soundness, privacy, eligibility, unreusability, fairness, and verifiability requirements of an election. As long as the two parties that made up the administrator did not cooperate to cheat during the election, the privacy of the election was not compromised by the administrator and the counter after the election even if they collaborated to do so. The distributed quantum election scheme also worked when the quantum channels contained losses and errors.

\section*{Acknowledgement}
This work was supported by the National Natural Science Foundation of China under Grant No.61173157.


\begin{thebibliography}{}
%\softraggedright
\itemsep=-4pt plus.2pt minus.2pt  %% sets the vertical space between items
\small
\bibitem{1}Chaum D 1988 {\textit Advances in Cryptology-Eurocrypt} 1988  (Berlin: Springer-Verlag) p177
\bibitem{2}Fujioka A, Okamoto T and Ohta K 1993 {\textit Lecture Notes in Computer Science} {\bf 718}
244
\bibitem{3}Cranor L F and Cytron R K 1996 Washington University Computer Science Technical Report 1996
\bibitem{4}Chaum D 1988 {\textit Journal of cryptology} {\bf 1}
65
\bibitem{5}Sako K and Killian J 1995 {\textit Advances in Cryptology-Crypto} 1994 (Berlin: Springer-Verlag) p411
\bibitem{6}Jafari S, Karimpour J, and Bagheri N 2011 {\textit International Journal on Computer Science and Engineering} {\bf 3}
2191
\bibitem{7}Cohen J D and Fischer M J 1985 {\textit 26th Annual Symposium on Foundations of Computer Science} p372
\bibitem{8}Chaum D 1981 {\textit Communications of the ACM} {\bf 24}
84
\bibitem{9}Marius I and Ionut P 2011 {\textit Computer Science Master Research} {\bf 1}
67
\bibitem{10}Ibrahim S, Kamat M, Salleh M, and Aziz S R A 2003 {\textit {$4^{th}$} National Conference on Telecommunication Technology Proceedings} p193
\bibitem{11}Ambainis A, Mosca M, Tapp A and Wolf R 2000 {Proceedings of the 41st Annual Symposium on Foundations of Computer Science}
p547, {\textit IEEE Computer Society Press} 2000
\bibitem{12}Boykin P and Roychowdhury V 2003 {\textit Phys. Rev.} A {\bf 67}
042317
\bibitem{13}Chen K and Lo H K 2005 arXiv:quant-ph/0404133
\bibitem{14}Deng F G, Li X H, Zhou H Y, and Zhang Z J 2005 {\textit Phys. Rev.} A {\bf 72}
044302
\bibitem{15}Wang T Y and Wen Q Y 2010 {\textit Chin. Phys.} B {\bf 19}
060307
\bibitem{16}Bennett C H and Brassard G 1984 {\textit Advances in Proceedings of the IEEE International Conference on Computers, Systems and Signal Processing}, India:Bangalore, December 10--12, 1984 p175
\bibitem{17}Ekert A K 1991 {\textit Phys. Rev.} Lett.  {\bf 67}
661
\bibitem{18}Bennett C H 1992 {\textit Phys. Rev.} Lett.  {\bf 68}
3121
\bibitem{19}Gisin N, Ribordy G, Tittel W, and Zbinden H 2002 {\textit Rev. Mod. Phys.} {\bf 74}
145
\bibitem{20}Christandl M and Wehner S 2005 {\textit Advances in Cryptology-Asiacrypt} 2005 {\bf 3788}
217
\bibitem{21}Vaccaro J A, Spring J, and Chefles A 2007 {\textit Phys. Rev.} A  {\bf 75}
012333
\bibitem{22}Hillery M, Ziman M, Buzek v, and Bielikov M 2006 {\textit Phys. Lett.} A {\bf 349}
75
\bibitem{23}Dolev S, Pitowsky I, and Tamir B 2006 arXiv:quant-ph/0602087
\bibitem{24}Horoshko D and Kilin S 2011 {\textit Phys. Lett.} A {\bf 375}
1172
\bibitem{25}Li Y and Zeng G 2009 {\textit Chinese Optics Letters} {\bf 7}
152
\bibitem{26}Okamoto T, Suzuki K, and Tokunaga Y 2008 {\textit NTT Thchnical Review} {\bf 6}
\bibitem{27}Zhou R R and Yang L 2012 {\textit Chin. Phys.} B {\bf 21}
8
\bibitem{28}Asharov G and Orlandi C 2012 {\textit Advances in Cryptology-Asiacrypt} 2012 p681
\bibitem{29}Durette B W 1999 {\textit Bachelor thesis, Massachusetts Institute of Technology}
\bibitem{30}Cramer R, Franklin M, Schoenmakers B and Yung M 1996 {\textit Advances in EUROCRYPT'96 Proceedings} (Berlin: Springer-Verlag) p72
\bibitem{31}Wiesner S 1983 {\textit ACM Sigact News} {\bf 15}
78
\bibitem{32}Yang L, Wu L A and Liu S H 2002 {\textit Acta Phys. Sin.} {\bf 51}
961(in Chinese)
\bibitem{33}Brassard G, Broadbent A, Fitzsimons J, Gambs S and Tapp A 2007 {\textit Advances in Cryptology-Asiacrypt} 2012 p460
\bibitem{34}Bouda J and Sprojcar J 2007 {\textit In Proceedings of the First
International Conference on Quantum, Nano, and Micro Technologies} 2007
%\bibitem{33}Li Y and Zeng G 2008 {\textit Optical Review} {\bf 15}
%219
%\bibitem{34}Bonanome M, Buzek V, Hillery M, and Ziman M 2011 {\textit Phys. Rev.} A {\bf 84}
%022331
\end{thebibliography}
\end{document}